\documentclass[twocolumn]{aastex631}
\usepackage{amsmath}

\shorttitle{}
\shortauthors{Isobe et al.}
\graphicspath{{./}{figures/}}
\begin{document}

%-- Title --%
%\title{ALMA ACA photometry at Band 8 
%to the west hot spot of the radio galaxy Pictor A.}
\title{
ALMA ACA detection of submillimeter emission   
associated with the west hot spot of the radio galaxy Pictor A}
%-- authors --%
% \correspondingauthor{Isobe Naoki}
% \email{n-isobe@ir.isas.jaxa.jp}
\author{Naoki Isobe}
\affiliation{
        Institute of Space and Astronautical Science (ISAS), 
        Japan Aerospace Exploration Agency (JAXA), 
        3-1-1 Yoshinodai, Chuo-ku, Sagamihara, Kanagawa, 252-5210, Japan}
\email{n-isobe@ir.isas.jaxa.jp}
%-
\author{Hiroshi Nagai}
\affiliation{
        National Astronomical Observatory of Japan,
       2-21-1 Osawa, Mitaka, Tokyo, 181-8588, Japan}
\affiliation{
    Department of Astronomical Science, 
    The Graduate University for Advanced Studies, 
    SOKENDAI, 2-21-1 Osawa, Mitaka, Tokyo 181-8588, Japan}
%-
\author{Motoki Kino}
\affiliation{
        Kogakuin University of Technology \& Engineering, 
        Academic Support Center,
        2665-1 Nakano, Hachioji, Tokyo, 192-0015, Japan}
\affiliation{
        National Astronomical Observatory of Japan,
       2-21-1 Osawa, Mitaka, Tokyo, 181-8588, Japan}
%-  
\author{Shunsuke Baba}
\affiliation{
	Graduate School of Science and Engineering, Kagoshima University, 
	1-21-35 Korimoto, Kagoshima, Kagoshima 890-0065, Japan}
\affiliation{
        National Astronomical Observatory of Japan,
       2-21-1 Osawa, Mitaka, Tokyo, 181-8588, Japan}	
%-
\author{Takao Nakagawa}
\affiliation{
       Institute of Space and Astronautical Science (ISAS), 
        Japan Aerospace Exploration Agency (JAXA), 
        3-1-1 Yoshinodai, Chuo-ku, Sagamihara, Kanagawa, 252-5210, Japan}s
%-
\author{Yuji Sunada}
\affiliation{
	Department of Physics, Saitama University, 
        255 Shimo-Okubo, Sakura-ku, Saitama, 338-8570, Japan}
%- 
\author{Makoto Tashiro}
\affiliation{
        Institute of Space and Astronautical Science (ISAS), 
        Japan Aerospace Exploration Agency (JAXA), 
        3-1-1 Yoshinodai, Chuo-ku, Sagamihara, Kanagawa, 252-5210, Japan}
\affiliation{
        Department of Physics, Saitama University, 
        255 Shimo-Okubo, Sakura-ku, Saitama, 338-8570, Japan}

%-- Abstract --%
\begin{abstract}
In order to investigate the far-infrared excess detected 
from the west hot spot of the radio galaxy Pictor A
with the Herschel observatory,
a submillimeter photometry is performed 
with the Atacama Compact Array (ACA) 
of the Atacama Large Millimeter/submillimeter Array 
at Band 8 with the reference frequency of 405 GHz.
A submillimeter source is discovered at the radio peak of the hot spot.
Because the 405 GHz flux density of the source, $80.7\pm3.1$ mJy, 
agrees with the extrapolation of the synchrotron radio spectrum, 
the far-infrared excess is suggested to exhibit no major contribution 
at the ACA band.
In contrast, by subtracting the power-law spectrum tightly constrained 
by the radio and ACA data,
the significance of the excess in the Herschel band is well confirmed. 
No diffuse submillimeter emission 
is detected within the ACA field of view, and thus, 
the excess is ascribed to the west hot spot itself. 
In comparison to the previous estimate based on the Herschel data,
the relative contribution of the far-infrared excess 
is reduced by a factor of $\sim 1.5$.
The spectrum of the excess below the far-infrared band is 
determined to be harder than that of the diffusive shock acceleration.
This strengthens the previous interpretation that 
the excess originates via the magnetic turbulence 
in the substructures within the hot spot.
The ACA data are utilized to evaluate  
the magnetic field strength of the excess 
and of diffuse radio structure associated to the hot spot.
\end{abstract}

\keywords{Radio hot spots (1344) --- Relativistic jets (1390)  --- Non-thermal radiation sources (1119) --- Fanaroff-Riley radio galaxies(526) --- Magnetic fields (994) --- Radio interferometers (1345)}

%===================================
\section{Introduction} 
\label{sec:intro}
%===================================
Fanaroff–Riley type-II \citep{fanaroff1974} radio galaxies 
generally host compact radio sources, called hot spots, 
at the terminal of their nuclear jets.  
In the standard picture, 
the hot spots are regarded as an ongoing site of 
the diffusive shock acceleration \citep{begelman1984}, 
where particles are energized up to relativistic regime. 
As a result, they are usually recognized  
as a strong synchrotron radio and inverse Compton X-ray emitter 
\citep{hardcastle2004}. 

%----------------------%
% Figure 1: ALMA image %
%----------------------%
\begin{figure*}[t]
\plotone{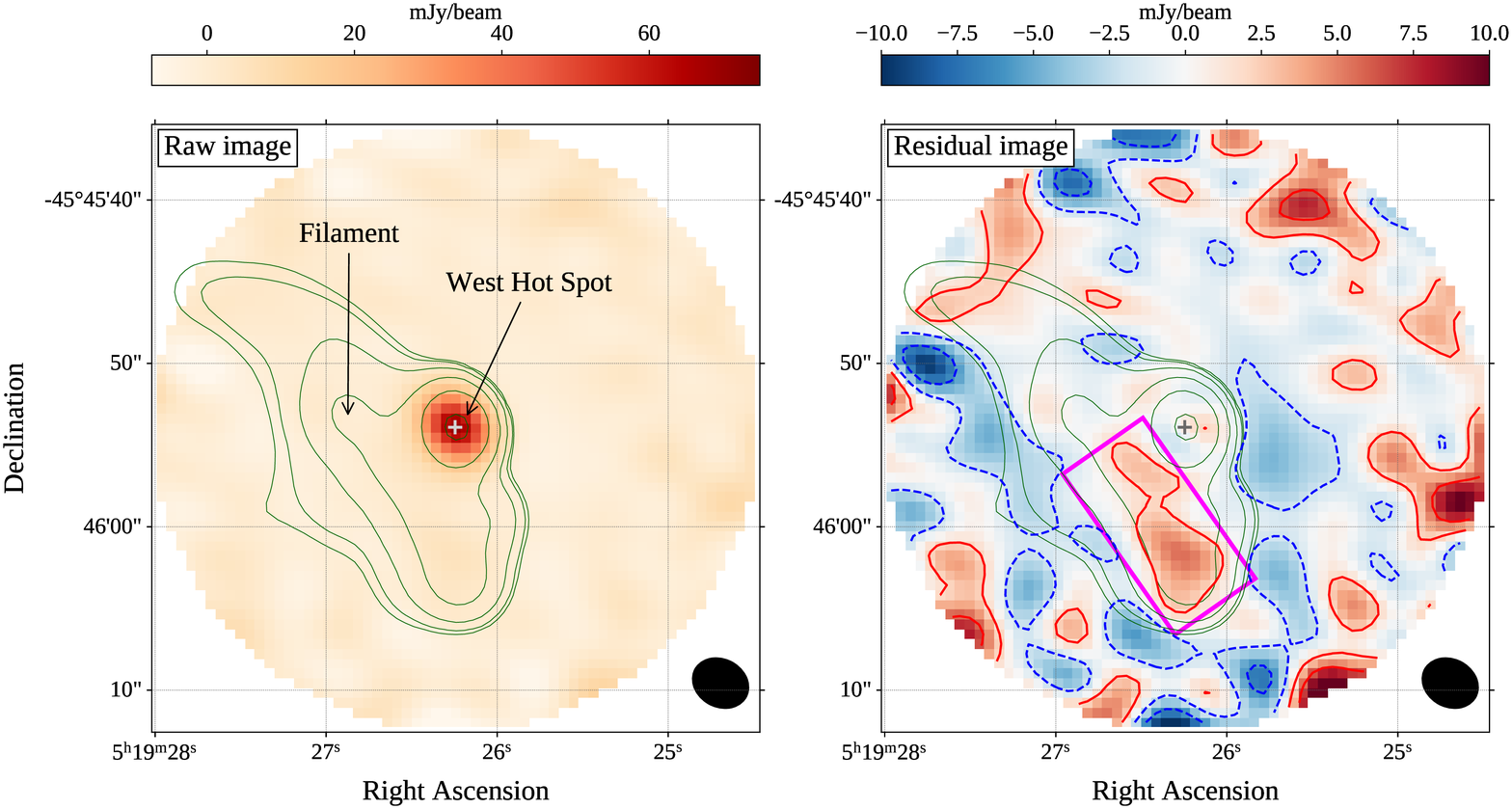}
\caption{
(left) 
Primary-beam-corrected ALMA ACA image at 405 GHz around the west hot spot of Pictor A,
on which the ATCA 4.86 GHz image 
\citep{isobe2017, isobe2020} 
is overlaid with thin solid contours. 
The arrows indicate the west hot spot and filament \citep{Roser1987}.
The ALMA ACA beam is shown with the ellipse on the bottom right. 
The cross points the ALMA ACA position of the detected submillimeter source.
(right) 
Residual image after subtracting the submillimeter source.
The negative and positive RMS noise levels 
($\pm 1 \sigma$ and $\pm 3 \sigma$)
are displayed with the dashed and solid contours, respectively.
The thick solid rectangle indicates the aperture  
adopted for evaluating the flux upper limit of the filament. }
\label{fig:image}
\end{figure*}

Theoretical and numerical studies \citep[e.g.,][]{inoue2009} indicate that magnetic turbulence is induced in the post-shock region.
The turbulence \citep[e.g.,][]{asano2014} and 
associated magnetic re-connection \citep[e.g.,][]{sironi2014} 
are theoretically proposed to behave as an efficient accelerator. 
However, concrete signatures of these acceleration processes 
in the hot spots have not yet been confirmed, 
except for an observational suggestion by \citet{orienti2017}.

These acceleration processes are predicted 
to generate a hard particle spectrum 
\citep[e.g.,][]{asano2014,sironi2014} 
with the corresponding synchrotron spectral index of $\alpha < 0.5$ 
where the flux density at the frequency $\nu$ 
is described as $F_\nu \propto \nu^{-\alpha}$.
This spectral slope is significantly flatter than 
that of the standard diffusive shock acceleration (i.e., $\alpha \ge 0.5$).
As a result, 
the turbulence and/or re-connection accelerations are possible 
to dominate the synchrotron spectrum in the higher frequency range,
namely in the submillimeter and/or infrared band,
even if their contribution is expected to be negligible 
in the radio band in comparison to the standard diffusive shock acceleration.

Recent progress in infrared investigation into hot spots 
of nearby radio galaxies \citep{isobe2017,isobe2020,sunada2022}
hinted a possible observational evidence 
of the turbulence and/or magnetic re-connection acceleration. 
By making most of the mid-infrared (MIR) data 
with the Wide-ﬁeld Infrared Survey Explorer (WISE) 
and the far-infrared (FIR) ones with 
the Spectral and Photometric Imaging REceiver  
\citep[SPIRE;][]{griffin2010} 
onboard the Herschel observatory,
\citet{isobe2017,isobe2020} have discovered 
from the west hot spot of the radio galaxy Pictor A
an excess emission in the FIR and MIR bands
over the main synchrotron component dominating the radio and optical data.
They successfully indicated that the excess exhibits 
a very hard spectrum with an index of $\alpha = 0.22 \pm 0.06$ 
below the FIR frequency range \citep{isobe2020}.
This excess was interpreted by invoking  
the turbulence and/or re-connection acceleration 
operated in 10-pc scale substructures  
which are resolved with the Very Long Baseline Array (VLBA) 
within the object \citep{tingay2008}.
However, the large SPIRE beam 
\citep[$17.6"$ for the PSW array at the wavelength of $250$ $\mu$m;][]{griffin2013}
made it difficult to resolve the west hot spot from neighboring diffuse structures 
including the so-called filament \citep{Roser1987},
and thus, 
prevented a definite identification of the excess with the hot spot.

% {\bf Importance of ALMA }
In order to tackle this difficulty,
an FIR or submillimeter instrument
equipped with an imaging capability 
in a spatial scale up to $\gtrsim 15"$ 
with no flux loss 
is desirable. 
The Atacama Compact Array \citep[ACA;][]{iguchi2009}
of the Atacama Large Millimeter/submillimeter Array (ALMA) 
is regarded as one of the ideal instruments. 
Especially, when it is operated at Band 8, 
the ALMA ACA simultaneously achieves a reasonable beam size ($\sim3"$)
and a large Maximum Recoverable Size (MRS; $\sim 15"$).
Therefore, an ALMA ACA Band-8 observation was conducted 
of the west hot spot of the radio galaxy Pictor A.

For the consistency to the previous studies \citep{tingay2008,isobe2017,isobe2020},
the following cosmological parameters are taken from \citet{spergel2003};
$H_0 = 71$ km s$^{-1}$ Mpc$^{-1}$, $\Omega_{\rm m} = 0.27$, 
and $\Omega_{\Lambda} = 0.73$. 
These give an angle-to-size conversion factor of $688$ pc arcsec$^{-1}$ 
at the redshift of Pictor A \citep[$z=0.035$;][]{eracleous2004}.

%===================================
% \newpage
\section{Observation} 
\label{sec:obs}
%===================================
The west hot spot of the radio galaxy Pictor A was targeted 
by the ALMA ACA at Band 8 on 2022 July 26,
as a project approved for the ACA supplemental call 
of the ALMA Cycle 8 2021 (the project code of 2021.2.00039.S). 
The nominal receiver and correlator configurations 
for continuum observations were adopted 
with the representative frequency of $405$ GHz. 
The peak position of the object, 
which is measured from the $4.86$ GHz radio image 
with the Australia Telescope Compact Array (ATCA)
% \footnote{kindly provided by Dr. Lenc; private communication}
as the right ascension and declination of 
$\alpha=05^{\rm h}19^{\rm m}26.2419^{\rm s}$ and 
$\delta=-45^{\rm d}45^{\rm m}53.971^{\rm s}$ respectively 
\citep{isobe2017}, was placed at the ACA field-of-view center. 
For the bandpass, phase and flux calibrations, 
the following three neighbouring radio sources were observed; 
J$0440-4333$, J$0515-4556$ and J$0538-4405$.

%===================================
\section{Analysis and Results} 
\label{sec:results}
%===================================
The standard pipeline products 
with the pipeline version of 2021.2.0.128 and 
the Common Astronomical Software Application \citep[CASA;][]{mcmullin2007} version of 6.2.1.7
were taken from the ALMA science archive.
No manual reprocessing was performed to the pipeline products.
The ALMA data were analyzed with the CASA version 6.5.1 
in combination with the python astronomical packages, 
{\tt Astropy} \citep{astropy2013,astropy2018,astropy2022} 
and {\tt photutils}. % \citep{photutils2022}.

The 405 GHz ALMA ACA interferometric image 
of the west hot spot of Pictor A 
is displayed in the left panel of Figure \ref{fig:image}. 
On the ALMA image, the 4.86 GHz ATCA contour image is superposed.
The figure clearly reveals a bright submillimeter source
spatially associated with the west hot spot.
A two-dimensional Gaussian image fitting procedure 
with the CASA tool {\tt imfit} was applied to 
the detected submillimeter source
with the sky level fixed at 0 mJy beam$^{-1}$.
The resultant parameters of the submillimeter source
are summarized in Table \ref{tab:photometry}.

%--------------------------% 
% Table 1: ALMA photometry %
% taken from photometry_2  %
% See page 18 in Note #1   %
%--------------------------%
\begin{deluxetable}{lll}
\tablecaption{Summary of the source parameters.}
\label{tab:photometry}
\tablecolumns{3}
\tablewidth{0pt}
%==================================================================
\tablehead{
	\colhead{Category} 	& \colhead{Parameter} 		& \colhead{Value}}
\startdata %------------------------------------------------------------------
Photometry & $F_\nu$(405 GHz) (mJy) \tablenotemark{a}	& $80.7 \pm 3.1$	\\
Astrometry & $\alpha$ (Right Ascension) 	    & $ 05^{\rm h}19^{\rm m}26.2458^{\rm s}$	\\
		& $\delta$ (Declination) 	& $-45^{\rm d}45^{\rm m}53.908^{\rm s}$ \\
		& ($\Delta \theta_{\alpha}$, $\Delta \theta_{\delta}$) \tablenotemark{b}	
							& $(0.035",   0.034")$	\\
Source shape \tablenotemark{c}	
		& $d_{\rm maj}$  \tablenotemark{d}	& $3.73" \pm 0.09"$\\
		& $d_{\rm min}$ \tablenotemark{e}	& $3.21" \pm 0.07"$\\
		& $\phi_{\rm PA}$ \tablenotemark{f}	& $46.8^\circ \pm 5.7^\circ$ \\
Beam shape	& $d_{\rm maj, b}$ \tablenotemark{d} 	& $3.55"$ \\
		& $d_{\rm min, b}$ \tablenotemark{e} 	& $2.92"$ \\
		& $\phi_{\rm PA, b}$ \tablenotemark{f}	& $61.5^\circ$ \\ 
\enddata %------------------------------------------------------------------
\tablenotetext{a}{Flux density at the frequency of 405 GHz.}
\tablenotetext{b}{The positional error along the right ascension and declination without any systematic errors included.}
\tablenotetext{c}{Before beam deconvolution.}
\tablenotetext{d}{The major size in the FWHM.}
\tablenotetext{e}{The minor size in the FWHM.}
\tablenotetext{f}{The position angle.}
% \tablenotetext{g}{The sky level is fixed at fixed at 0 mJy beam$^{-1}$.}
\end{deluxetable}

The peak position measured with the ALMA ACA 
(the cross in Figure \ref{fig:image}) 
is found to agree with the ATCA and WISE positions
\citep{isobe2017},
within $\sim 75$ mas and $\sim 310$ mas, respectively. 
With the typical ALMA astrometric accuracy 
($\sim$ 5\% of the synthesized beam, 
corresponding to $\sim 150$ mas in this ACA image) taken into account, 
these positional coincidences naturally 
indicate a physical connection between the ATCA, ALMA and WISE sources.
The beam-inclusive major and minor sizes of the source, 
$d_{\rm maj} = 3.73" \pm 0.09"$ and $d_{\rm min} = 3.21" \pm 0.07"$ 
respectively in the full width at half maximum (FWHM), 
are slightly larger than the synthesized ALMA ACA beam size.
This suggests the intrinsic FWHM source size of $\sim 1"$ 
after the beam deconvolution. 
The integrated 405 GHz flux density of the source 
was measured as $F_{\nu} = 80.7 \pm 3.1$ mJy.

The right panel of Figure \ref{fig:image} shows 
the residual image derived with the {\tt imfit} tool
after the Gaussian profile of the detected source was subtracted. 
Within the annulus with a radius of $r=6"$--$15"$ centered 
on the detected source, 
where the impact from the source was negligible, 
the root-mean-squared (RMS) noise level was 
estimated as $\sigma = 2.02$ mJy beam$^{-1}$.
In the residual image, 
the $\pm 1 \sigma$ and $\pm 3 \sigma$ noise levels 
are drawn with the solid (positive) and dashed (negative) lines. 
Except for the region near the edge of the field of view 
(e.g., $r \gtrsim 15"$) 
where spatial fluctuations are enhanced by the primary-beam correction, 
no region was found to exhibit 
a significance higher/lower than $\pm 3 \sigma$.

The residual image hints possible emission 
associated with the south part of the filament,
although its significance is relatively low. 
Thus, the upper limit on its flux density 
was estimated from the aperture shown with the thick solid rectangle 
in the right panel of Figure \ref{fig:image}.
The aperture has a length and width of $12"$ and $6"$,
corresponding to the physical size of $8.26$ kpc and $4.13$ kpc 
respectively, at the source frame.
By supposing a spatial independence at a scale larger than the beam size,
the RMS noise level was scaled by the square root of the beam number 
within the aperture. 
As a result, the ALMA ACA upper-limit flux density was derived 
as $F_{\nu,{\rm fil}} = 15.0$ mJy 
at the signal-to-noise (SN) ratio of $3$.

%=========================
\section{Revisit to the FIR excess of the west hot spot} 
\label{sec:FIRexcess}
%=========================
\subsection{Significance}
\label{sec:significance}
\begin{figure*}[htbp]
\plotone{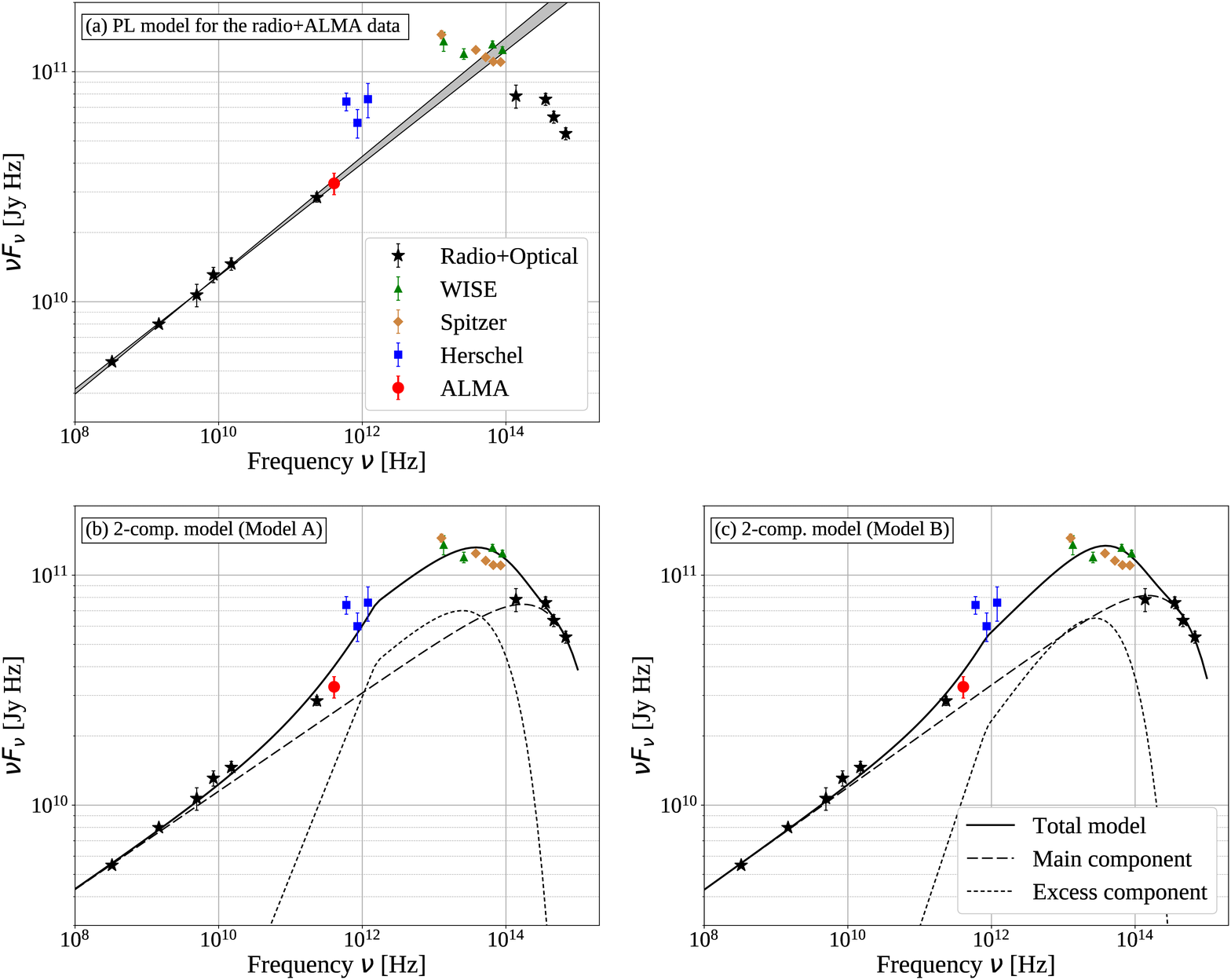}
\caption{Synchrotron SEDs of the west hot spot of Pictor A. 
         The ALMA data point is shown with the circle, 
         the optical and radio data \citep{meisenheimer1997} 
         are plotted with the stars, 
         the FIR data with Herschel \citep{isobe2020} %
         are displayed with the squares, and the MIR data with WISE and Spitzer 
         \citep[][and reference therein]{isobe2017} 
         are shown with the triangles and diamonds, respectively.     
         In panel panel (a), the best-fit PL model to the radio and ALMA data
         is shown with the gray area encompassed 
         with the two solid lines (the 68\% confidence region). 
         In panels (b) and (c), the two-component model originally proposed 
         in \citet{isobe2020} is plotted with the thick solid line, 
         with the dashed and dotted lines 
         indicating the main CPL and excess BPL components, respectively.
         The spectral parameters are taken from \citet{isobe2020} (Model A in Table \ref{tab:SED}) in panel (b),
         while all the the parameters are tuned by including the ALMA data 
         (Model B) in panel (c).
         }
\label{fig:sed}
\end{figure*}

%=========================
The spectral energy distribution (SED) of the synchrotron emission 
from the west hot spot of Pictor A
in the radio-to-optical range is shown in Figure \ref{fig:sed}.
The ALMA data on this plot (the filled circle) takes into account 
the systematic error of the ALMA photometry at Band 8 
(i.e., typically $10$\%
\footnote{Taken from the ALMA Cycle 8 2021 Technical Handbook}; 
$\Delta F_{\rm \nu} = 8.1$ mJy in this case)
by a root sum square.
The figure clearly depicts that 
the ALMA flux is consistent with the simple extrapolation of the radio spectrum.
The best-fit power-law (PL) model to the radio and ALMA data
is plotted with the gray area 
encompassed by the solid lines (the $68$\% confidence range)
in panel (a) of Figure \ref{fig:sed}.
The model yields the $5$ GHz flux density and energy index of 
$F_{\nu}(\rm{5~GHz}) = 2.18 \pm 0.03$ Jy 
and $\alpha = 0.748 \pm 0.007$, respectively.
These parameters agree 
with those derived without the ALMA data \citep{meisenheimer1997}.
This result indicates that the FIR excess 
does not have a major contribution 
in the ALMA frequency range.

By simply extrapolating the PL model best fit to the radio and ALMA data,
the FIR excess revealed with the Herschel SPIRE \citep{isobe2020}
is evaluated.
This method is expected to provide 
a conservative estimate on the significance of the FIR excess,  
since the observed high-frequency radio and submillimeter flux is
possibly contaminated by the excess flux (see \S\ref{sec:SED}).  
Panel (a) of Figure \ref{fig:sed} displays that the PL extrapolation 
(the gray area, $F_{\nu, {\rm PL}}$ in Table \ref{tab:FIRexcess}) 
is insufficient to describe the SPIRE flux. 
The signal statistics of the FIR excess ($F_{\nu, {\rm ex}}$)
in the three SPIRE photometric bands 
(i.e., the PLW, PMW and PSW)
are listed in Table \ref{tab:FIRexcess}.
Even though adopting this conservative estimate,
the significance of the FIR excess is fairly confirmed. 

Thanks to its large MRS ($16.5"$ in the FWHM at 405 GHz),
the ALMA ACA image shown in Figure \ref{fig:image} covers 
with almost no flux loss
the SPIRE beam at least for the PSW array
\citep[$17.6"$ in the FWHM;][]{griffin2013}.
Importantly, the ALMA ACA data did not reveal 
a diffuse source 
with a spatial scale comparable to the SPIRE beam. 

A possible contamination from the filament to the FIR excess 
is evaluated,
by combining the radio and ALMA data
(see \S \ref{sec:filament} for the details).
When the broken-PL (BPL) model 
which connects the radio data in the GHz range and the ALMA upper limit 
is simply extrapolated to the higher frequency range,
the upper limit on the SPIRE flux of the filament 
($F_{\nu, {\rm fil}}$)
is evaluated as tabulated in Table \ref{tab:FIRexcess}
(the arrowed squares in Figure \ref{fig:sed_fil}).
It is found that the contribution from the filament is 
less than $\lesssim24$\%, and thus,
a large fraction of the FIR excess is ascribed to the west hot spot itself.

%----------%
% Figure 2 %
%----------%

%---------------------% 
% Table 2: FIR excess %
%---------------------%
\begin{deluxetable}{llll}
\tablecaption{Summary of the FIR excess.}
\label{tab:FIRexcess}
\tablecolumns{4}
\tablewidth{0pt}
%==================================================================
\tablehead{
	\colhead{SPIRE bands} 	& \colhead{PLW} & \colhead{PMW} & \colhead{PSW}
}
\startdata %-----------------------------------------------------------------
$\nu$ ($10^{12}$ Hz)        &  $0.60$             
                            &  $0.86$                  
                            &  $1.20$ \\
%---
$F_\nu$ (mJy)\tablenotemark{a}              
                            & $123.7 \pm 10.9$              
                            & $ 70.0 \pm  9.9$                      
                            & $ 63.3 \pm 10.7$  \\
%---
$F_{\nu, {\rm PL}}$ (mJy)\tablenotemark{b}              
                            & $60.6 \pm 1.9$             
                            & $46.4 \pm 1.6$                      
                            & $36.1 \pm 1.3$  \\
%---
$F_{\nu,{\rm ex}}$ (mJy)\tablenotemark{c}               
                            & $63.0 \pm 11.1$              
                            & $23.6 \pm 10.1 $                       
                            & $27.2 \pm 10.8 $  \\
SN ratio                    & $5.70 $              
                            & $2.34 $                      
                            & $2.52 $  \\
\hline
% $F_{\nu,{\rm fil}}$ (mJy)\tablenotemark{d}               
%                            & $ \le 10.1$              
%                            & $ \le 7.0 $                       
%                            & $ \le 5.0 $  \\ % ATCA-ALMA PL
$F_{\nu,{\rm fil}}$ (mJy)\tablenotemark{d}               
                            & $ \le 8.9$              
                            & $ \le 5.6$                       
                            & $ \le 3.6$  \\ % VLA-ALMA BPL
\enddata %------------------------------------------------------------------
\tablenotetext{a}{The total FIR flux of the west hot spot 
                    taken from \citet{isobe2020}}
\tablenotetext{b}{Simple extrapolation of the PL model 
                    reproducing the radio and ALMA data.}
\tablenotetext{c}{Excess flux over the PL model.}
\tablenotetext{d}{Upper limit ($ {\rm SN} =3$) on the filament flux density.}
\end{deluxetable}

%----------------------% 
% Table 3: SED fitting %
%----------------------%
\begin{deluxetable*}{llll}
\tablecaption{Summary of the SED parameters.}
\label{tab:SED}
\tablecolumns{4}
\tablewidth{0pt}
%==================================================================
\tablehead{
	\colhead{Component} 	& 
	\colhead{Parameters} 	& 
	\colhead{Model A\tablenotemark{a}}        & 
	\colhead{Model B\tablenotemark{b}}
}
\startdata %------------------------------------------------------------------
Main CPL    & $F_{\nu, {\rm m}}(\rm 5~GHz)$ (Jy)  
                & $1.98 \pm 0.09 $    
%               & $2.18$ (fix)  \\
                & $2.05 \pm 0.11$\\
            & $\alpha_{\rm m}$          
                & $0.79 \pm 0.02$
%               & $0.748$ (fix)   \\
                & $0.78 \pm 0.02$ \\
            & $\nu_{\rm c,m}$ (Hz)      
                & $8.0_{-2.6}^{+3.9} \times 10^{14}$  
%               & $4.5_{-2.9}^{+3.1} \times 10^{14}$  \\
                & $6.8_{-2.1}^{+3.0} \times 10^{14}$ \\
Excess BPL 
        & $F_{\nu,{\rm e}}(\rm 1.67~GHz)$ (mJy) 
                & $121.3$ (fix) 
%               & \\
                & $38_{-34}^{+281}$\\
%        & $F_\nu(\rm 405~GHz)$ (mJy)
%               & $36.3$
%               & $12.0^{+12.8}_{-6.2}$    \\
%               & --- \\
        & $\alpha_{\rm low,e}$ 
                & $0.22 \pm 0.06 $ 
%               & $-0.09\pm0.66$  \\
                & $0.06 \pm 0.35 $ \\
        & $\alpha_{\rm high,e}$ \tablenotemark{c} 
                & \multicolumn{2}{c}{$\alpha_{\rm low,e} + 0.5 $ } \\
        & $\nu_{\rm b,e}$ (Hz)      
                & $1.6_{-1.0}^{+3.0}\times 10^{12}$
%               & $8.6\times10^{11}$ ($<1.2\times10^{14}$) \\
                & $8.6_{-7.8}^{+90.4} \times 10^{11}$\\ 
        & $\nu_{\rm c,e}$ (Hz)      
                & $8.8_{-2.4}^{+3.4} \times 10^{13}$ 
%               & $3.8^{+3.7}_{-1.9} \times 10^{13}$ \\
                & $6.3_{-3.0}^{+5.8} \times 10^{13}$ \\
\enddata %------------------------------------------------------------------
\tablenotetext{a}{Case 2 in \citet{isobe2020}}
\tablenotetext{b}{All the parameters are revised with the ALMA data included.}
\tablenotetext{c}{The standard cooling break condition \citep{carilli1991} is assumed.}
\end{deluxetable*}

%=========================
\subsection{Spectral interpretation}
\label{sec:SED}
%=========================
In order to investigate the FIR excess, 
\citet{isobe2020} adopted a simple two-component model,
where the main cutoff-PL (CPL) component describes the radio and optical data, 
and the BPL one subjected to the high-frequency cutoff 
is adopted to reproduce the excess. 
Because the BPL component was suggested to require 
a low-frequency spectral slope significantly flatter than 
that of the diffusive shock acceleration,
they invoked for the FIR excess 
the turbulence acceleration \citep[e.g.,][]{asano2014} 
and/or magnetic re-connection one \citep[e.g.,][]{sironi2014}.
In addition, the substructures inside the object
resolved with the VLBA \citep{tingay2008} 
were proposed as the acceleration site, 
since the sum flux of the substructures at 1.67 GHz
was found to agree with the BPL flux. 

Table \ref{tab:SED} tabulates 
eight controlling parameters of the two-component model;
the flux density at a given frequency ($F_{\nu, {\rm m}}$),
spectral index ($\alpha_{\rm m}$), 
and cutoff frequency ($\nu_{\rm c,m}$)
of the main CPL component,
and the flux density ($F_{\nu,{\rm e}}$), 
spectral indices below and above the break 
($\alpha_{\rm low,e}$ and $\alpha_{\rm high,e}$),
the break and cutoff frequencies 
($\nu_{\rm b,e}$ and $\nu_{\rm c,e}$)
of the excess BPL component.
The standard cooling break under the continuous energy injection condition
\citep[$\Delta \alpha = \alpha_{\rm high,e} - \alpha_{\rm low,e} =0.5$;]
[]{carilli1991} is adopted for the BPL component.
The result in \citet{isobe2020} is referred to 
as Model A in Table \ref{tab:SED},
and the corresponding model curve 
is shown in panel (b) of Figure \ref{fig:sed}.
The ALMA data does not necessarily reject Model A,
because their deviation is estimated 
as a significance of only ${\rm SN}\sim 2.1$.

By including the ALMA data, 
the spectral parameters are tuned for a better solution.  
The best-fit model is displayed in panel (c) of Figure \ref{fig:sed},
and the corresponding SED parameters 
are listed in Table \ref{tab:SED} (Model B).
While most parameters are found to remain relatively unchanged from Model A,
the relative contribution of the excess in the FIR-to-MIR band
is reduced by a factor of $\sim 1.5$,
because the ALMA data, 
consistent with the simple radio-PL extrapolation, 
gives a tight constraint. 
The $1.67$ GHz flux density of the BPL component for the FIR excess,
$F_{\nu,e}(1.67~\rm{GHz}) = 38_{-34}^{+281}$ mJy in Model B, 
is still consistent to the sum flux density of the substructures 
\citep[$121.3$ mJy;][]{tingay2008}
though with a large error in $F_{\nu,e}(1.67~\rm{GHz})$. 
Especially, the low-frequency spectral index of the excess BPL component, revised as $\alpha_{\rm low,e} = 0.06 \pm 0.35$,
is confirmed to be significantly harder than 
that of the diffusive shock acceleration 
($\alpha = 0.5$ for the strong shock limit). 
Instead, this value agrees with the prediction
from the turbulence acceleration 
\citep[$\alpha = 0$;][]{asano2014}
under the hard-sphere condition \citep{park1995}.
Additionally,  
the intrinsic size of the ALMA source, $\sim 1"$,
is comparable to the distance among the substructures. 
These results strengthen the scenario  
that the FIR excess is produced through the turbulence acceleration 
operated in the VLBA substructures \citep{isobe2020}.

%=========================
\subsection{physical condition}
\label{sec:physical_condition}
%=========================
The magnetic field for the FIR excess of the west hot spot 
is evaluated after \citet{isobe2020}.
When the observed break is ascribed to the radiative cooling,
the break frequency is converted into the magnetic field 
as
\begin{equation}
B^3 \simeq
\frac{27 \pi e m_{\rm e} v^2  c}{\sigma_{\rm T}^2} L_{\rm cool}^{-2} \nu_{\rm b}^{-1},
\label{eq:Bme}
\end{equation} 
where $e$ is the electron charge, $m_{\rm e}$ is the electron rest mass,
$\sigma_{\rm T}$ is the Thomson cross section, 
$v$ is the down flow speed in the shock frame, $c$ is the speed of light, and $L_{\rm cool}$ is the cooling length during which 
the impact of the radiative cooling on the electrons 
at the break becomes effective \citep{inoue1996}.

Here, the VLBA substructures inside the object 
are regarded as the origin of the FIR excess 
after \citet{isobe2017,isobe2020}.
The minor length of the individual substructures \citep[$28$--$58$ pc;][]{tingay2008}
is adopted as the cooling length,
since their minor axis is nearly aligned to the jet 
toward the west hot spot. 
The flow speed is assumed as $v\sim0.3c$ \citep{kino2004}.
Thus, the cooling break frequency of the FIR excess,
$\nu_{\rm b,e} = 8.6_{-7.8}^{+90.4} \times 10^{11}$ Hz (Model B),
is transformed to the magnetic field as $B=0.86$--$7.2$ mG
by Equation (\ref{eq:Bme}).

Alternatively, 
the magnetic field under the minimum-energy condition 
is widely adopted as a standard indicator
for synchrotron sources \citep{miley1980}. 
The shape of the emission region is assumed to be a simple disk, 
of which the height and diameter, respectively,
equal to the minor and major lengths of the substructures \citep[$87$--$170$ pc for the latter;][]{tingay2008}.
The lowest synchrotron frequency 
of $\nu_{\rm min} = 10$ MHz is adopted after \citet{miley1980},
and the proton contribution is simply neglected.
As a result, the determined spectrum of the FIR excess 
roughly yields the minimum-energy magnetic field 
as $B_{\rm me} = 280$--$480$ $\mu$G.
Importantly, the magnetic field from the cooling break 
is found to be higher than the minimum-energy value
as $B/B_{\rm me} = 2$--$26$.

Several ideas are briefly discussed to reconcile this discrepancy. 
The above $B_{\rm me}$ value is possibly underestimated 
due to undetectable plasma contents.
Firstly, the contribution from low-energy electrons is considered. 
The existence of the low-energy electrons 
with a Lorentz factor of $\gamma_{\rm e} \ll 100$ 
associated to the jets and hot spots 
is under debate \citep[e.g.,][]{brunetti2001}.
However, owing to the spectral hardness of the FIR excess,
the minimum-energy magnetic field is found to stay unchanged,
even if the minimum synchrotron frequency is 
lowered down to $\nu_{\rm min}=2$ kHz, 
which corresponds to $\gamma_{\rm e} \sim 1$ for 
the inferred $B_{\rm me}$ value.

The second alternative is the proton contribution. 
The above calculation conservatively assumes 
the proton-to-electron energy ratio of $k=0$,
because the proton is invisible through the synchrotron emission.
If the protons have a significant energy dominance of $k\sim1000$,
the minimum-energy magnetic field is enhanced 
up to $B_{\rm me} =2$--$3.5$ mG. 
Thus, the observed cooling break becomes 
consistent to the minimum-energy condition.
However, such a high proton dominance is possibly doubtful 
in Fanaroff–Riley type-II sources \citep[e.g.,][]{kawakatsu2016,kino2012}.

Finally, a physical mechanism to strengthen the magnetic field 
is invoked to interpret the inconsistency 
between the cooling break and minimum-energy condition.
It is theoretically predicted that 
the post-shock turbulence is possible to 
amplify the magnetic field by more than a factor of 10
\citep[e.g.,][]{inoue2009,mizuno2011}. 
This mechanism is successfully applied to X-ray shells, 
rims and hot spots found in Galactic supernovae remnants 
\citep[e.g.,][]{bamba2003,uchiyama2007}.
As is indicated in \citet{isobe2020},
the magnetic amplification by the post-shock turbulence is 
well matched to the value of $B/B_{\rm me} = 2$--$26$ 
obtained by including the ALMA data.
This idea also seems compatible with the interpretation 
that the FIR excess is produced via the turbulence acceleration.

%=========================
\subsection{additional notes}
\label{sec:notes}
%=========================
Before closing the discussion on the properties of the FIR excess,
this subsection briefly presents several concerns
to be settled through future observations.
In spite of their frequency proximity,
the total flux density of the west hot spot 
measured with the SPIRE PLW at 600 GHz (see Table \ref{tab:FIRexcess})
is by a factor of $\sim 1.5$ higher than 
that with the ALMA ACA at 405 GHz (tabulated in Table \ref{tab:photometry}).
Accordingly, the spectrum in this narrow range 
appears to be steeply ``rising" 
with a corresponding two-point spectral index of $\alpha \sim -1.0$.
Contamination from point sources is probably rejected
as discussed in \S \ref{sec:results},
since no significant source is detected 
on the ALMA ACA image shown in Figure \ref{fig:image},
of which the field of view is comparable to the SPIRE PLW beam
\citep[the FWHM size of $35.2"$;][]{griffin2013}.
However, strictly speaking, this PLW beam 
is not fully covered with the MRS of the ALMA ACA at 405 GHz.
Then, it is potential for the PLW flux density 
to have been affected by unidentified diffuse emission 
with a scale of $\gtrsim 16.5"$.
In contrast, as mentioned in \S \ref{sec:significance},
the higher-frequency SPIRE results 
are basically regarded as more reliable, 
since the SPIRE beam at these bands is comparable to 
the ALMA ACA MRS.

The substructures are previously resolved 
only in the 1.67 GHz VLBA observation \citep{tingay2008}.
Correspondingly, there is a very wide frequency gap 
in more than 2 orders of magnitude between the substructures and FIR excess. 
In addition, 
due to the low declination of Pictor A ($\delta\sim-46^\circ$),
the VLBA observation was conducted 
in a relatively severe observational condition 
with only 5 southern VLBA antennas 
at a low elevation angle (namely $\lesssim 13^\circ$).
For better understanding the substructures,
it is important to obtain high-resolution images 
with a higher dynamic range in improved observational conditions. 
Such observations are expected to be enabled 
by future very long baseline interferometric observatories,
especially in the southern hemisphere.

Finally, 
there is possible room for improvement in the spectral modeling.
Panel (c) of Figure \ref{fig:sed} suggests that
the BPL model (Model B in Table \ref{tab:SED}) appears 
to underestimate slightly the FIR flux density,
although the deviation between the modeled and measured
flux densities is insignificant.
A current lack in spatially resolved spectral data 
(e.g., sufficiently better than $\sim 1"$)
especially over the submillimeter, FIR and MIR ranges 
is thought to have made it difficult to determine 
precisely the spectral shape of the FIR excess.

By making most of ALMA,
most of the above issues are probably cleared up.
ALMA observations with an angular resolution of $\lesssim 100$ mas
enables to identify the submillimeter counterpart to the VLBA substructures. 
The Square Kilometer Array is similarly thought to be useful 
to derive a radio insight into the substructures.
In addition, a combination of the Near Infrared Camera and 
Mid-Infrared Instrument %\citep{rieke2015}
onboard the James Webb Space Telescope is expected 
to give a great progress in understanding the FIR excess.

%----------%
% Figure 3 %
%----------%
\begin{figure}[t]
\plotone{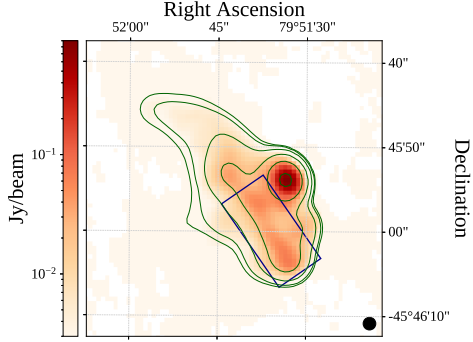}
    \caption{
    The 8.44 GHz VLA image of the west hot spot of Pictor A \citep{perley1997}, 
    displayed after the astrometric correction.
    The 4.86 GHz ATCA image, 
    shown in Figure \ref{fig:image},
    is superposed with the thin contours. 
    The photometric aperture for the filament is plotted with 
    the thick solid rectangle. 
    The filled circle on the bottom right indicates 
    the VLA beam with the FWHM size of $1.5"$.
    }
    \label{fig:img_fil}
\end{figure}

%=========================
\section{Properties of the filament}
\label{sec:filament}
%=========================
% The VLA data of the west hot spot of Pictor A, 
% provided by Perley \citep{perley1997},
% are analysed.

The ALMA ACA data is also useful to infer 
the physical condition in the filament.
The relativistic plasma filling the filament 
is thought to be supplied from the west hot spot 
through back or lateral flows \citep{mizuta2010,saxton2002}. 
Thus, immediately after the plasma is injected into the filament,
its synchrotron spectral index in the GHz range is 
simply supposed to be same as 
that of the main component in the spectrum of the west hot spot 
($\alpha_{\rm m} = 0.78 \pm 0.02$ for Model B).
The radio data obtained with the Very Large Array (VLA)
basically support this idea,
since the spectral index of the filament in the $5$--$15$ GHz range
measured with VLA \citep[$\alpha \sim 0.8$;][]{perley1997}
is consistent to the index of the main component.
In contrast, electrons in the filament radiating the synchrotron emission 
at the ALMA band are expected to be already subjected 
to the radiative cooling,
since the ALMA flux of the filament is insignificant, 
as shown in \S \ref{sec:results}.
These indicate that the filament exhibits 
the cooling break between the VLA and ALMA frequency ranges.

A combination of the VLA and ALMA data are utilised 
to quantitatively evaluate the cooling break frequency of the filament.
Here, the VLA images at the frequency of 4.87, 8.44, and 14.96 GHz 
with a beam size of $1.5"$ are adopted from \citet{perley1997}.
The lower-frequency VLA data presented in \citet{perley1997},
i.e. in the range of $74$ MHz -- 1.4 GHz, are not employed 
in the following investigation, 
since their beam size is found to be insufficient 
to resolve the filament from the west hot spot.
It is noticed that the adopted VLA images are affected by 
an astrometric error in an angular scale of $\lesssim 0.7"$. 
Thus, the sky coordinates of the VLA images are corrected 
by referring to the peak position of the west hot spot 
on the 4.86 ATCA image in Figure \ref{fig:image},
because the astrometry of this ATCA image was 
previously validated within $\sim 20$ mas in \citet{isobe2017}.
Figure \ref{fig:img_fil} presents
the 8.44 GHz VLA image around the west hot spot after the astrometric correction,
in comparison to the ATCA image.
The ATCA image is omitted in the following photometric analysis
due to its larger beam size ($\sim 2.5"$),
although it was checked that the 4.86 GHz ATCA flux density of the filament 
agrees with the 4.87 GHz VLA one within $\sim 3$\%.

\begin{figure}[t]
\plotone{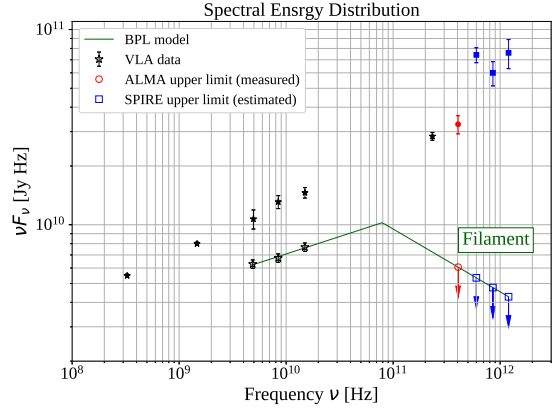}
    \caption{
    The synchrotron SED of the filament.
    The open stars show the VLA data, 
    while the arrowed open circle indicates the ALMA upper limit.
    The BPL model connecting the VLA and ALMA data is 
    drawn with the solid line,
    where the spectral index below the break is determined 
    from the PL fitting to the VLA data 
    ($\alpha_{\rm fil} = 0.82 \pm 0.06$)
    and the standard cooling break condition is assumed 
    ($\Delta \alpha = 0.5$).
    The upper limit on the SPIRE flux, 
    evaluated from the simple BPL extrapolation,
    is plotted with the arrowed open squares.  
    For comparison, the spectrum of the west hot spot is 
    displayed by adopting the same symbol notation
    as for Figure \ref{fig:sed}.}
    \label{fig:sed_fil}
\end{figure}

The VLA flux density of the filament is integrated 
within the same photometric aperture as for the ALMA image,
which is replotted in Figure \ref{fig:img_fil} 
with the thick solid rectangle.
The 8.44 GHz flux density of the filament is measured  
as $F_{\nu, {\rm fil}}= 0.80 \pm 0.04$ Jy with VLA.
Here, the typical systematic error of the VLA photometric calibration
\citep[$5$\%;][]{perley2017} is conservatively adopted. 
The SED of the filament is displayed in Figure \ref{fig:sed_fil}.
The obtained VLA spectrum,
shown with the open stars in Figure \ref{fig:sed_fil},
is well reproduced by a single PL model.
The spectral index, derived as $\alpha_{\rm fil} = 0.82 \pm 0.06$,
is confirmed to agree with the result reported in \citet{perley1997}. 

The ALMA upper limit on the flux density of the filament,
which is plotted with the open circle 
accompanied with the downward arrow in Figure \ref{fig:sed_fil},
is found to be significantly below the simple extrapolation of 
the PL model to the VLA data.
Thus, it is demonstrated that the cooling break of the filament 
is located between the VLA and ALMA bands.
The PL model to the VLA data is connected to the ALMA upper limit
by introducing a simple break (i.e., the BPL model), 
as drawn with the solid line in Figure \ref{fig:sed_fil}.
Here, the standard cooling break condition 
of $\Delta \alpha = 0.5$ is employed. 
As a result, the cooling break frequency of the filament is 
evaluated as $\nu_{\rm b, fil} \lesssim 8 \times 10^{10}$ Hz.

%----------%
% Figure 4 %
%----------%

By utilizing equation (\ref{eq:Bme}),
the magnetic field of the filament is calculated 
from the break frequency.
For the evaluation, 
two representative cases are considered for the cooling length;
the distance between the hot spot and filament along the jet 
($L_{\rm cool,1}\sim3.4$ kpc or $5"$)
and 
the width of the filament perpendicular to the jet 
toward the filament aperture for the flux evaluation 
($L_{\rm cool,2}\sim6.2$ kpc or $9"$).
In either case, the back (for $L_{\rm cool,1}$) or lateral (for $L_{\rm cool,2}$)
flow speed of $v\sim0.3c$ \citep{perley1997,kino2004} is adopted. 
The magnetic field strength in the filament is evaluated 
from the derived value of $\nu_{\rm b,fil}$ as 
$B_1 \gtrsim 280$ $\mu$G and $ B_2 \gtrsim 190$ $\mu$G
for $L_{\rm cool,1}$ and $L_{\rm cool,2}$, respectively.
The magnetic field of the filament is suggested to be 
by a factor of $\sim 3$--$40$ weaker 
than that of the FIR excess from the west hot spot.

By utilizing the BPL spectrum connecting the VLA and ALMA data
and adopting the rectangular aperture 
in Figures \ref{fig:image} (b) and \ref{fig:img_fil},
a crude estimated on the minimum-energy magnetic field of 
the filament is given as $B_{\rm me} \sim 30 $ $\mu$G.
Similar to the FIR excess, the filament is found to exhibit 
a high value of $B/B_{\rm me} \sim 5$--$10$.
These indicate that the turbulence also plays 
an important role in the filament. 

\begin{acknowledgments}
The authors are grateful to the anonymous reviewer for suggestive comments.
The ALMA data with the following code are utilized in the present study; 
{\tt ADS/JAO.ALMA\#2021.2.00039.S}. 
ALMA is a partnership of ESO (representing its member states), NSF (USA) and NINS (Japan), together with NRC (Canada), MOST and ASIAA (Taiwan), and KASI (Republic of Korea), in cooperation with the Republic of Chile. The Joint ALMA Observatory is operated by ESO, AUI/NRAO and NAOJ.
The VLA images of the west hot spot of Pictor A 
in the electric form are kindly provided by Dr. R. A. Perley.
The ATCA data of Pictor A were originally 
provided by Dr. E. Lenc at the time of \citet{isobe2017}.
This research is supported by the JSPS KAKENHI grants 
No. JP18K03709, JP21H01137, JP21H04496, JP21K03635, and JP22H00157. 
This research is aided 
by the ALMA Japan Research Grant of NAOJ ALMA Project, NAOJ-ALMA-276. 
\end{acknowledgments}

\software{
CASA \citep{mcmullin2007},
Astropy \citep{astropy2013,astropy2018,astropy2022},
photutils %\citep{photutils2022}
}

%--------------%
% bibliography %
%--------------%
\bibliographystyle{aasjournal}
\bibliography{ms}

\end{document}